\documentclass{ws-procs975x65}

\begin{document}

\title{NEGATIVE KOMAR MASSES IN REGULAR STATIONARY SPACETIMES}

\author{MARCUS ANSORG}

\address{Max-Planck-Institut f\"ur Gravitationsphysik,
	Albert-Einstein-Institut,\\
	D-14476 Golm, Germany\\
	\email{mans@aei.mpg.de}}

\author{DAVID PETROFF}

\address{Theoretisch-Physikalisches Institut, University of Jena,
	Max-Wien-Platz 1, \\ D-07743 Jena, Germany\\
	\email{D.Petroff@tpi.uni-jena.de}}

\begin{abstract}
A highly accurate multi-domain spectral method is used to study axially symmetric and stationary spacetimes containing a black hole or disc of dust surrounded by a ring of matter. It is shown that the matter ring can affect the properties of the central object drastically. In particular, by virtue of the ring's frame dragging, the so-called Komar mass of the black hole or disc can become negative. A continuous transition from such discs to such black holes can be found.
\end{abstract}

\bodymatter
\vspace*{1cm}

We study self-gravitating systems in equilibrium, consisting of a uniformly rotating, homogeneous
perfect fluid ring surrounding a central object which is either a black hole or a rigidly rotating disc of dust. The corresponding space-time is characterized by two Killing vectors $\eta$ and $\xi$ which describe axisymmetry and stationarity respectively. For such configurations, Bardeen \cite{Bardeen73} assigns to each of the two objects a mass based on the Komar integral \cite{Komar59}. In a similar way, he assigns an angular momemtum to each of the objects.
This definition of mass is applicable both to matter and black holes and can be used for single components
of a many-body problem in stationary spacetimes. We will refer to it here as the Komar mass even when applied to single
objects.

If one deals with the Komar mass, a natural questions concerns its positive definiteness. We address this question by analyzing the following formulae

\begin{equation}\label{Smarr}
	M_{\rm h} = \frac{\kappa A}{4\pi} + 2\Omega_{\rm h} J_{\rm h}
\end{equation}
and 
\begin{equation}\label{Smarr_disc}
	 M_{\rm d} = e^{V_0^{\rm d}}M_0 + 2\Omega_{\rm d} J_{\rm d},
\end{equation}
valid for central black hole and central disc configurations respectively.

In Eqn. (\ref{Smarr}), the central black hole's Komar mass $M_{\rm h}$ is related to (i) its surface gravity $\kappa$, (ii) its horizon area $A$, (iii) the angular velocity $\Omega_{\rm h}$ of the horizon and (iv) the black hole's (Komar) angular momentum $J_{\rm h}$. For single black holes Eqn. (\ref{Smarr}) was given by Smarr \cite{Smarr73} but it holds true even in the presence of a surrounding ring \cite{Bardeen73} (see also Ref. \cite{Carter73}).

A similar expression can be derived for rigidly rotating discs of dust with and without a surrounding ring of matter (cf.\ III.15 in Ref. \cite{BW71}). In Eqn. (\ref{Smarr_disc}) the disc's Komar mass $M_{\rm d}$ is given in terms of (i) a constant $e^{V_0^{\rm d}}$ that is related to the relative redshift $Z_0^{\rm d}$ of photons emitted from the center of the disc and observed at infinity ($Z_0^{\rm d} = e^{-V_0^{\rm d}} - 1$), (ii)
the baryonic mass $M_0$ of the central disc, (iii) its angular velocity $\Omega_{\rm d}$ and (iv) its (Komar) angular momentum $J_{\rm d}$.

The first summands on the right hand sides of formulae (\ref{Smarr}) and (\ref{Smarr_disc}) are always positive. However, either of these terms can become small if we assume the horizon area and the baryonic mass to be finite and consider the central object to be close to a degenerate black hole. In particular we find a continuous transition from the central disc to the central black hole configurations \cite{AnsorgPetroff06}, and at the transition point the central object is a degenerate black hole for which the first terms in Eqns. (\ref{Smarr}, \ref{Smarr_disc}) vanish.

For the discussion of the sign of the second summands, a `frame dragging'-effect of the central object caused by the surrounding ring is important. If the torus is highly relativistic and quickly rotating, it creates a large ergosphere (a portion of space in which the Killingvector $\xi$ is spacelike). In this case, a counter-rotating central object  (i.e.~the sign of its angular momentum is opposite to that of the torus) inside the ergosphere is dragged along the direction of the motion of the ring's fluid elements. As a consequence, the corresponding angular velocity of the central object can assume the same sign as that of the surrounding ring, and thus the second summand becomes negative. 

Combining the two arguments, it is possible to identify negative Komar masses by considering central objects close to a degenerate black hole and counter-rotating with respect to the torus. Note that only highly relativistic and quickly rotating tori will exert a sufficiently large frame dragging effect to bring this about. Moreover, the specific rate of counter-rotation must be limited since very strong counter-rotation would lead to opposite signs of the two angular velocities, $\Omega_{\rm h/d}$ and $\Omega_{\rm ring}$, and hence to a positive second summand.

In Ref. \cite{AnsorgPetroff06} we construct sequences of both central black hole and central disc configurations along which the Komar mass of the central object becomes negative. In addition, we show that the Komar mass can become negative on either side of the continuous parametric transition from matter to a black hole.

The results presented reveal clearly that the Komar mass is not an intrinsic property of a gravitational source but rather a feature of an object within a specific highly relativistic spacetime geometry. The interesting question
about the maximally attainable ratio
\begin{equation}\label{Ratio}
-M^{\mathrm {negative}}/M^{\mathrm {positive}},
\end{equation}
where $M^{\mathrm {negative}}$ and $M^{\mathrm {positive}}$ are the sums of all negative and positive Komar mass components, respectively, in a given stationary space-time, will be the subject of a future publication.\footnote{Note that by virtue of the positive mass theorem (see e.~g.~Ref. \cite{Gibbons-et-al83} and references therein), the ratio (\ref{Ratio}) is always less than 1 for regular physically relevant space-times obeying the dominant energy condition.}

\vfill
%\pagebreak

\end{document}